\documentclass[twocolumn,showpacs,preprintnumbers,floatfix,amsmath,amssymb]{revtex4}

\usepackage{dcolumn}
\usepackage{bm}
\usepackage{graphicx}
\usepackage[pdfpagemode=None,pdfstartview=FitH,pdftitle={Argon spectral function and neutrino interactions},pdfauthor={A.M. Ankowski and J. T. Sobczyk},bookmarks=true]{hyperref}

\newcommand{\ve}[1]{\ensuremath{\mathbf{#1}}}
\DeclareMathOperator{\Tr}{\ensuremath{\text{Tr}}}
\newcommand{\sla}[1]{\ensuremath{\slash\mspace{-9mu}#1}}
\newcommand{\n}[1]{\ensuremath{|\mathbf{#1}|}}
\newcommand{\EB}{\ensuremath{\overline\epsilon_B}}

\begin{document}
\title{Argon spectral function and neutrino interactions}

\author{Artur M. Ankowski}
\author{Jan T. Sobczyk}
\email{jsobczyk@ift.uni.wroc.pl}

\affiliation{Institute of Theoretical Physics\\ University of
Wroc{\l}aw,\\ pl. Maksa Borna 9, 50-204 Wroc{\l}aw, Poland}

\date{\today}

\begin{abstract}
The argon spectral function is constructed and applied to
neutrino-argon cross section computations in the plane wave impulse
approximation with the Pauli blocking final state interaction effect
taken into account. The approximations of the construction method
are critically analyzed using the example of oxygen for which more
detailed computations are available. An effective description of
nucleus based on the information contained in a spectral function is
proposed. It is demonstrated that its predictions are close to those
obtained from the complete spectral function. The effective
description can be easily applied in Monte Carlo event generators.
\end{abstract}

\pacs{13.15.+g, 25.30.Pt}

\maketitle
\section{Introduction}
In recent years, a lot of interest has been concentrated on
improving the description of neutrino interactions~\cite{ref:NuInt}.
The motivations behind this research are future precise neutrino
experiments which require better knowledge about the cross sections.
It also makes crucial whether the theoretical models can be
implemented in Monte Carlo (MC) event generators.

Many papers focus on neutrino energies of a~few GeV because this is
the proposed energy range  of several long-baseline experiments. In
this energy region one has to consider both quasielastic and
inelastic charge current (CC) processes. It is common to factorize
neutrino-nucleus interaction into two steps: an interaction with a
\emph{single} bound nucleon (impulse approximation) and then
reinteractions inside nucleus. The main difficulty is caused by the
need of precise evaluation of the nuclear effects.

Existing MC codes rely on various versions of the Fermi gas (FG)
model because it is straightforward to implement them in the MC
routine. The simplest FG model with two adjustable
parameters---Fermi momentum and average binding
energy~\cite{ref:Smith&Moniz}---can be modified by inclusion of the
recoil nucleus energy in the kinematics~\cite{ref:Bodek&Ritchie} and
by local density effects \cite{ref:LDA}. It is, however, clear from
the experience gained in describing the electron-nucleus
interactions that the model can be considered only as the first
approximation and that the MC simulations based on the FG model can
be neither detailed nor precise.

The approach which introduces more realistic picture of a~nucleus is
the one using the spectral function (SF)~\cite{ref:Frullani&Mougey}.
The SF describes a~distribution of momenta and removal energies of
nucleons inside nucleus. In theoretical models, the SF combines
contributions from the shell model (the mean field part) and from
the short-range correlations. The second one (about 20\% of the
overall strength) accounts for the higher momentum and removal
energy
values~\cite{ref:Ciofi&Simula,ref:Benhar&Fabrocini&Fantoni&Sick}.
Recently, this part has been directly observed
experimentally~\cite{ref:Rohe}. Computations based on the SF
reproduce precisely the electron scattering data in the region of
the quasielastic peak~\cite{ref:Benhar&Farina&Nakamura}.

In this paper we adopt the plane wave impulse approximation (PWIA)
and neglect the effects of final state interactions (FSI) except
those from Pauli blocking. Evaluation of FSI  is necessary in order
to compare the predictions of the model with experimental data. It
is argued in Refs.~\cite{ref:Benhar&Farina&Nakamura,ref:FSI} that in
computation of the inclusive cross section (after summing over the
final hadronic states) the construction of the SF and the treatment
of FSI are two independent problems. Calculation of FSI effects is
usually based on the complex optical potential. In the case of
electron scattering, FSI do not change much the overall cross
section at the fixed angle: the elastic peak is quenched and the
cross section is redistributed making the peak wider. For the GeV
neutrino interactions, i.e. when most of the quasielastic cross
section corresponds to small values of $Q^2$, it is useful to
distinguish Pauli blocking from the remaining FSI effects, because
impact of the latter on the cross section is quite small (compare,
for example fig.~14 in~\cite{ref:Benhar&Farina&Nakamura}).

There is a lot of discussion in the literature on the FSI effects in
other approaches to quasielastic neutrino-nucleus scattering. In
some of the recent papers~(e.g.,
Ref.~\cite{ref:Leitner&Alvarez-Ruso&Mosel}), FSI is implied to be
responsible only for a redistribution of the inclusive cross section
among various exclusive observable channels. A similar meaning of
FSI is adopted in many MC generators of events \cite{ref:NuInt}.
However in other theoretical approaches, the FSI effects modify also
the inclusive cross section. In~\cite{ref:Nieves}, it is achieved by
using the dressed nucleon propagator with the contribution from the
self-energy. References~\cite{ref:Maieron} and~\cite{ref:Meucci}
describe the final nucleon wave function as a solution of the Dirac
equation with the relativistic optical potential, the inclusive
cross section is then obtained if the imaginary part of the
potential is neglected. For $1$~GeV neutrinos the FSI effects are
reported to reduce the inclusive cross section by $\sim10\%$.

We confine our consideration to quasielastic reactions (the primary
vertex), but we would like to stress that its extension to the
resonance region is straightforward. Several papers consider the two
dynamics together~\cite{ref:Leitner&Alvarez-Ruso&Mosel,ref:Marteau},
also in the formalism of the spectral
function~\cite{ref:Nakamura&Seki}. (Recent discussion of the
resonance region can be found
in~\cite{ref:Sato,ref:Lalakulich&Paschos,ref:Benhar&Meloni}.)
However, we focus on the construction of the spectral function, and
the inclusion of the resonance region is not likely to provide an
extra input. In the discussion of the resonance region some new
theoretical uncertainties appear: the properties of resonances in
the nuclear matter and  the nonresonant
background~\cite{ref:ResCompl}. It is well known that the
theoretical understanding of the basic dynamics is still not
complete since in the case of electron scattering it is difficult to
reproduce the cross section in the dip region between the elastic
and $\Delta$ resonance peaks \cite{ref:dip}.

For a~few-GeV neutrino one has to consider also more inelastic
channels. They are described by means of the DIS formalism with the
specific techniques to include the nuclear effects. A recent
comprehensive study discussed the following nuclear effects: Fermi
motion and nuclear binding (FMB), off-shell corrections (OS),
nuclear pion excess and coherent processes~\cite{ref:Kulagin&Petti}.
The results of the analysis show that in the kinematical region of
large Bjorken $x$ variable only FMB and OS effects are relevant.
Both FMB and OS corrections are present in our approach: FMB is the
core of the spectral function computations and OS were included in
the de Forest prescription of evaluating weak current hadronic
matrix elements~\cite{ref:de Forest}.

There has been a lot of effort to calculate the spectral
functions~\cite{ref:Ciofi&Simula,ref:Benhar&Fabrocini&Fantoni&Sick,ref:Benhar&Fabrocini&Fantoni}.
Systematic computations exist only for light nuclei up to oxygen and
also for infinite nuclear matter. The aim of this paper is to
calculate the spectral function for argon and to discuss its
application to computation of the neutrino-argon cross section.
Liquid-argon technology has been successively tested in the case of
the ICARUS T600 module~\cite{ref:ICARUS} and is now considered as an
option in many neutrino experiments \cite{ref:Argon}. One of the
experiments is the T2K where liquid-Ar detector is planned to be an
element of the 2~km detector complex~\cite{ref:T2K}.

In our computations, we follow the approach proposed by Kulagin and
Petti~\cite{ref:Kulagin&Petti} which is partially based on the paper
of Ciofi degli Atti and Simula~\cite{ref:Ciofi&Simula}. The
computations contain several approximations, and it is necessary to
understand how precise the results are. For this reason we first
analyze the approximation in the case of oxygen, the heaviest
nucleus for which detailed computations of the spectral function
exist. Our conclusion is that the agreement is satisfactory, and we
pursue the computations for the argon.

Our paper is organized as follows. In Sec.~\ref{sec:CrossSection} we
introduce basic definitions and formulas, Sec.~\ref{sec:SM} shows
how the spectral function corresponds to the Fermi gas model and in
Sec.~\ref{sec:SpectralFunction} we present the approach of Kulagin
and Petti~\cite{ref:Kulagin&Petti}. In Sec.~\ref{sec:Oxygen}, we
discuss in detail the predictions of the spectral function
constructed for oxygen and compare them with the results obtained by
Benhar {\it et al.}~\cite{ref:Benhar&Farina&Nakamura}. Most
important are results contained in Secs.
\ref{sec:Argon}--\ref{sec:EffectiveDescription}. In
Sec.~\ref{sec:Argon} we construct the argon spectral function. Then,
having in mind implementations in MC event generators, in
Sec.~\ref{sec:EffectiveDescription} we investigate a~procedure to
extract an essential information contained in the spectral function.
The information is encoded in two functions. The first one is the
distribution of the nucleon momenta, and the second is the average
binding energy for a given value of momentum. These two functions
can be inserted in the Smith-Moniz formula for the inclusive cross
section which can be easily implemented in a~MC routine. Sec.
\ref{sec:Con} contains the conclusions.

\section{Spectral Function in Plane Wave Impulse Approximation}\label{sec:PWIA}

\subsection{Neutrino-nucleus inclusive cross section}\label{sec:CrossSection}

We denote the initial (final) lepton four-momentum as $k$ ($k'$) and
the hadronic initial (final) four-momentum as $p$ ($p^f$). In
general, several hadrons or nuclei fragments can be in the final
state $|f(p^f)\rangle$. The four-momentum transfer is $q\equiv
(\omega,\ve q)\equiv k-k'$.

The neutrino-nucleus cross section is calculated in the Fermi theory
approximation (in the neutrino energy range of a few GeV the
condition  $|q^2|\ll M_W^2$ is satisfied).

The inclusive neutrino-nucleus differential cross section reads:
\begin{equation}
\frac{d\sigma}{d^3k'}= \frac{G_F^2\cos^2\theta_c}{8\pi^2E_\nu E_\mu}L_{\mu\nu}W^{\mu\nu},%
\end{equation}
where the leptonic tensor is defined as
\begin{eqnarray}
L_{\mu\nu}&\equiv&\frac14\Tr[\gamma_\mu(1-\gamma_5)(\sla{k}+m_l)\gamma_\nu(1-\gamma_5)(\sla{k'}+m_l')]\nonumber\\%
&=&2(k_\mu k'_\nu+k'_\mu k_\nu-k\cdot k'\thinspace g_{\mu\nu}-i\epsilon_{\mu\nu\rho\sigma}k^\rho k'^\sigma).%
\end{eqnarray}
and the nuclear tensor is
\begin{eqnarray}\label{eq:W^munu}
W^{\mu\nu}&\equiv&
\overline{\sum_{f,\,i}}\langle f(p^f)|{\cal J}^\mu(0)|i(M_A)\rangle\langle f(p^f)|{\cal J}^\nu(0)|i(M_A)\rangle^*\nonumber\\
&&\times(2\pi)^6\delta^4(q+p_A-p^f),
\end{eqnarray}

It is convenient to express the cross section in terms of energy
and momentum transfer:
\begin{equation}
\frac{ d\sigma}{ d\omega d\n{q}}=\frac{G^2_F\cos^2\theta_C}{4\pi}\frac{\n q}{E_\nu^2} L_{\mu\nu} W^{\mu\nu}.
\end{equation}

In this paper we adopt PWIA, i.e. the approximation in which
neutrino interacts with separate nucleons and the nucleon produced
in the interaction leaves the nucleus without further
reinteractions. We assume that
\[
|f(p^f)\rangle = |R(p_R)\rangle\otimes |p'\rangle .
\]

The spectral function $P(\ve p, E)$ is defined as the probability
distribution to remove from the nucleus a nucleon with momentum $\ve
p$ and leaving the residual nucleus with energy $E_R=M_A-M+E$:
\begin{eqnarray}
P(\ve p, E)&\equiv&\overline{\sum_{R,\,i}}\:\delta(M_A-E_R-M+E)\nonumber\\
&&\times\big|\langle R(p_R)|a(\ve p)|i(M_{A})\rangle\big|^2.%
\end{eqnarray}
In our notation, $|i(M_{A})\rangle$ is the state of the initial
nucleus (assumed to be at rest) of mass $M_A$, $|R(p_R)\rangle$ is
the final nucleus state (after a nucleon of momentum $\ve p'$ was
ejected) of four-momentum $p_R=(E_R, \ve p_R)$, and $M$ denotes the
nucleon mass. Summation over all the final states and averaging over
the spin states of the initial nucleus is performed. We omitted spin
and isospin dependencies of the annihilation operator, but one has
to keep in mind that there are two separate spectral functions: one
for protons and another for neutrons.

The spectral function contains a lot of information about the
nucleus. It follows from its definition that
\begin{eqnarray}\label{eq:momentumDistrib}
n(\ve p)&\equiv&\int P(\ve p, E)\,d E=\overline{\sum_{R,\,i}}\:\big|\langle R(p_R)|a(\ve p)|i(M_{A})\rangle\big|^2\nonumber\\%
&=& \overline{\sum_{i}}\langle i(M_{A})|a^{\dagger}(\ve p)a(\ve p)|i(M_{A})\rangle%
\end{eqnarray}
is the distribution of momenta inside the nucleus.

We normalize  the nucleon distribution of momenta $n(\ve p)$ (and
then also the spectral function) as
\begin{equation}\label{eq:normalization}
\int d^3p\:n(\ve p) =\int d^3p\,dE P(\ve p, E)=\begin{cases}
Z&\text{for protons,}\\
N &\text{for neutrons.}
\end{cases}
\end{equation}

In PWIA it is straightforward  to obtain the following expression
for the inclusive cross section:
\begin{eqnarray}
\frac{d\sigma}{d\omega d\n q}&=&\frac{G^2_F\cos^2\theta_C}{4\pi}\frac{\n q}{E_\nu^2}
\int dE\:d^3p\:\frac{P(\ve p, E)}{E_{\ve p}E_{\ve {p'}}}\nonumber\\
&&\times \delta(\omega+M-E-E_\ve{p'})L_{\mu\nu}H^{\mu\nu}.
\end{eqnarray}
where
\begin{eqnarray}\label{eq:H^munu}
H^{\mu\nu}&=&(2\pi)^6{\overline{\sum_\text{spins}}}{E_\ve{p}}{E_\ve{p'}} \langle\ve{p'}|\mathcal J^\mu(0)|\ve{p}\rangle\langle\ve{p'}|\mathcal J^\nu(0)|\ve{p}\rangle^* \nonumber\\%
&=&\frac{M^2}2\Tr\Big(\Gamma^\mu\frac{\sla p+ M}{2M}\gamma_0\Gamma^{\nu\dagger}\gamma_0\frac{\sla{p'}+M}{2M}\Big)%
\end{eqnarray}
and
\begin{eqnarray}
\Gamma^\mu&=&\gamma^\mu[F_1(q^2)+F_2(q^2)]-\frac{(p+p')^\mu F_2(q^2)}{2M}\nonumber\\
&&+\gamma^\mu\gamma_{5}F_A(q^2)+\gamma_{5}\frac{q^\mu F_P(q^2)}M.%
\end{eqnarray}

The hadronic tensor $H^{\mu\nu}$ has to be further specified because
nucleons described by~\eqref{eq:H^munu} are off shell. We will adopt
the standard de Forest prescription \cite{ref:de Forest}: to
approximate off-shell kinematics one can use free spinors and free
form factors, taking into account that only a~part of the energy is
transferred to the interacting nucleon, and the rest is absorbed by
the spectator system. It means that one makes a substitution
\begin{equation}
q=(\omega,\ve q) \rightarrow \widetilde q\equiv(\widetilde\omega,\ve
q)
\end{equation}
where $\widetilde\omega =\omega-\EB$, so that
\begin{equation}
H^{\mu\nu} \rightarrow \widetilde H^{\mu\nu}
\end{equation}

The model is finally specified by the form of the hadronic tensor:
\begin{eqnarray}
\frac{H^{\mu\nu}}{M^2}&=&-g^{\mu\nu}H_1+\frac{p^\mu p^\nu}{M^2}H_2+i\varepsilon^{\mu\nu\kappa\lambda}\frac{p_\kappa q_\lambda}{2M^2}H_3\nonumber\\%
&&-\frac{q^\mu q^\nu}{M^2}H_4+\frac{p^\mu q^\nu +q^\mu p^\nu}{2M^2}H_5
\end{eqnarray}
where
\begin{eqnarray}
H_1&=&F_A^2(1+\tau)+\tau (F_1+F_2)^2,\nonumber\\[5pt]
H_2&=&F_A^2+F_1^2+\tau F_2^2,\nonumber\\[5pt]
H_3&=&2 F_A(F_1+F_2),\nonumber\\
H_4&=&\frac14 F_2^2(1-\tau)+\frac12 F_1 F_2+F_A F_P-\tau F_P^2\nonumber\\
H_5&=&H_2,
\end{eqnarray}
with the notation $\tau=-q^2/(4M^2)$.

\subsection{Fermi Gas model}\label{sec:SM}
In the FG model
\begin{equation}
P(\ve p, E)\propto \theta(p_F-|\ve p|)
\end{equation}
and one only has to decide about the way in which the energy is
balanced.

In the previous section we used the energy conservation expressed as
[compare the $\delta$ function in Eq.~\eqref{eq:W^munu}]:
\begin{equation}
M_A+\omega=E_\ve{p'}+\sqrt{(M_{A-1})^2+\ve p^2}.
\end{equation}
We recall that the initial nucleus state is assumed to be factorized
into the neutron with momentum~$\ve p$ and the spectator system with
momentum~$-\ve p$. The neutron energy can be written as
\[
\sqrt{M^2+\ve p^2}-\epsilon_B(\ve p)\equiv E_{\ve p}-\epsilon_B(\ve p)%
\]
with binding energy~$\epsilon_B(\ve p)\geq0$, so that
\begin{equation}
M_A=\sqrt{(M_{A-1})^2+\ve p^2}+E_{\ve p}-\epsilon_B(\ve p)
\end{equation}
and the energy conservation takes form
\begin{equation}\label{ref:EB(p)}
E_{\ve p}-\epsilon_B(\ve p)+\omega=E_\ve{p'}.
\end{equation}

The Smith--Moniz approach to the FG model~\cite{ref:Smith&Moniz} is
to approximate~$\epsilon_B(\ve p)$ by the constant average
value~$\EB$, that is,
\begin{equation}\label{ref:EB}
E_{\ve p}-\EB+\omega=E_\ve{p'}.
\end{equation}
The cross section calculated within this model (with Pauli blocking
taken into account) is equal to
\begin{eqnarray}
\frac{d\sigma_\text{S-M}}{d E_\mu}&=&\frac{G_F^2\cos^2\theta_C}{4\pi
E_\nu^2}\frac{3N}{4\pi p_F^3}
\int d \n{q}\, d^3p\:\theta(p_F-{\n p})\nonumber\\%
&&\times\delta\big(\omega+E_{\ve p}-\EB-E_\ve{p'}\big)\:\theta(\n{p+q}- p_F)\nonumber\\%
&&\times\frac{\n q}{E_{\ve p}E_\ve{p'}}L_{\mu\nu}\widetilde H^{\mu\nu}.%
\end{eqnarray}

Using the concept of the spectral function one can say that the
Smith-Moniz~\cite{ref:Smith&Moniz} version of the FG model is
defined as
\begin{eqnarray}\label{eq:SM}
P_\text{S-M}(\ve p, E)&=&{\cal N}\delta(\sqrt{M^2+\ve p^2}-\EB-M+E)\nonumber\\%
&&\times\theta(p_F-|\ve p|)
\end{eqnarray}
where $\EB$ is an average binding energy.

In the Bodek-Ritchie version of the FG
model~\cite{ref:Bodek&Ritchie}, one avoids the approximations in the
energy delta function, obtaining
\begin{eqnarray}\label{eq:BR}
P_\text{B-R}(\ve p, E)&=&{\cal N}\delta (M_A-\sqrt{(M_{A-1})^2+\ve p^2}-M+E)\nonumber\\
&&\times\theta(p_F-|\ve p|).%
\end{eqnarray}
In Ref.~\cite{ref:Bodek&Ritchie}, the quadratic momentum
distribution is modified by adding higher momentum tail.  ${\cal N}$
is a~normalization constant (the same in both cases).

In~\eqref{eq:SM} and~\eqref{eq:BR} Pauli blocking (PB) was ignored,
because the usual PB term present in the expression for the cross
section [i.e. $\theta (|\ve p+\ve q|-p_F)$] depends on the momentum
transfer and therefore is not defined merely by characteristics of
the nucleus. It is natural to treat it separately as a FSI effect.

\subsection{Spectral function}\label{sec:SpectralFunction}

In theoretical models, the SF consists of isoscalar and isovector
parts~\cite{ref:Kulagin&Petti}:
\begin{equation}
P(\ve p,E)=P_\text{sc}(\ve p,E)+P_\text{vec}(\ve p,E).
\end{equation}

It is
common~\cite{ref:Benhar&Farina&Nakamura,ref:Kulagin&Petti,ref:Ciofi&Liuti&Simula}
to think of the isoscalar SF as itself consisting of two parts: the
mean field (MF) and the short-range correlation parts, that is,
\begin{equation}\label{eq:isoscalarSF}
P_\text{sc}(\ve p,E)=\frac{N+Z}2\left[P_\text{MF}(\ve p,E)+P_\text{corr}(\ve p,E)\right],%
\end{equation}
with the normalization given by the condition
\begin{equation}
\int dE~d^3p\:\left[P_\text{MF}(\ve p,E)+P_\text{corr}(\ve p,E)\right]=1.%
\end{equation}

\begin{figure}
\graphicspath{{plot/}}
\includegraphics[width=0.48\textwidth]{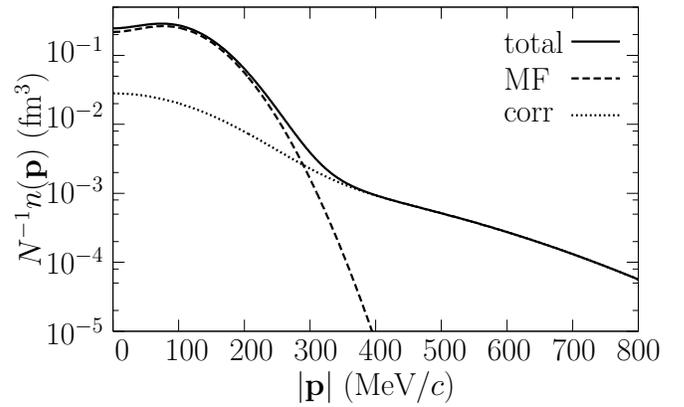}%
\caption{\label{fig:nContribs} Momentum distribution of nucleons
in $^{16}$O~\cite{ref:Ciofi&Simula}. Dashed and dotted line shows
the mean field and correlation contribution, respectively. The
solid line represents their sum.}
\end{figure}

As a consequence, the momentum
distribution~\eqref{eq:momentumDistrib} can also be represented as
the sum of two contributions:
\begin{equation}
n(\ve p)=\frac{N+Z}2\left[n_\text{MF}(\ve p)+n_\text{corr}(\ve p)\right].%
\end{equation}
Both $n_\text{MF}(\ve p)$ and $n_\text{corr}(\ve p)$ for a~few
nuclei are given in~\cite{ref:Ciofi&Simula}. In
Fig.~\ref{fig:nContribs} we see how they contribute to the overall
momentum distribution of $^{16}$O. These data will be used in the
next section.

The mean field part $P_\text{MF}(\ve p,E)$ provides proper
description of low-energy and low-momentum nucleons: they behave as
Fermi gas in self-consistent mean field potential with energy levels
given by the shell model. The presence of the correlation term
$P_\text{corr}(\ve p,E)$ reflects the fact that in the ground state
$NN$-correlations generate high momenta and high removal energies.

The low-energy contribution can be expressed
as~\cite{ref:Ciofi&Simula&Frankfurt&Strikman}
\begin{equation}
P_\text{MF}(\ve p,E)=\frac1A\sum_{\alpha<\alpha_F}A_{\alpha}n_{\alpha}(\ve p)\delta\big(E_{\alpha}+E_R(\ve p)-E\big)%
\end{equation}
with the recoil energy of the residual nucleus $E_R(\ve p)={\ve
p^2}/{(2M_{A-1})}$. In above equation $A=N+Z$ is the total number of
nucleons, and $A_\alpha$ number of nucleons in the shell-model
state~$\alpha$ of energy~$E_\alpha$ and momentum density
distribution $n_{\alpha}(\ve p)$ given by momentum space wave
functions.  In our convention, the values of $E_\alpha$ are positive
below the Fermi level $\alpha_F$.

When we replace $E_{\alpha}$ by the separation energy $E^{(1)}$
averaged over single-particle nucleon levels, the $\delta$ function
can be taken outside the sum and the expression simplifies
to~\cite{ref:Kulagin&Petti}
\begin{eqnarray}\label{eq:meanFieldSF}
P_\text{MF}(\ve p,E)&=&\delta\big(E^{(1)}+E_R(\ve p)-E\big)%
\frac1A\sum_{\alpha<\alpha_F}A_{\alpha}n_{\alpha}(\ve p)\nonumber\\%
&=&n_\text{MF}(\ve p)\,\delta\big(E^{(1)}+E_R(\ve p)-E\big).%
\end{eqnarray}

From $(e,e'p)$ experiments we know that~\eqref{eq:meanFieldSF}
gives, in fact, only a~part of $P(\ve p,E)$ because the depletion of
states is observed of a typical value $\sim 0.2$. It means that
$\sim 20\%$ of the strength of the spectral function must be
represented by a physics outside the shell model. It is usually
explained in terms of short-range correlations producing pairs of
high-momentum nucleons. To model $P_\text{corr}(\ve p,E)$, one
assumes that two nucleons form a~cluster with high relative momentum
and other $(A-2)$ nucleons remain
soft~\cite{ref:Ciofi&Liuti&Simula}. Within this picture the
following formula can be obtained~\cite{ref:Kulagin&Petti}:
\begin{eqnarray}\label{eq:correlationSF}
P_\text{corr}(\ve p,E)&=&n_\text{corr}(\ve p)\frac{M}{\n p}\sqrt{\frac\alpha\pi}\nonumber\\%
&&\times\left[\exp(-\alpha \ve p_\text{min}^2)-\exp(-\alpha\ve p_\text{max}^2)\right],\qquad%
\end{eqnarray}
where $\alpha=3/(4\langle \ve p^2\rangle \beta)$ with the mean
square of momentum $\langle \ve p^2\rangle$, $\beta=(A-2)/(A-1)$,
and
\begin{equation}\begin{split}
{\ve p}_\text{min}^2&=\left[\beta \n p - \sqrt{\smash[b]{2M\beta[E-E^{(2)}-E_R(\ve p)]}}\,\right]^2,\\%
{\ve p}_\text{max}^2&=\left[\beta \n p + \sqrt{\smash[b]{2M\beta[E-E^{(2)}-E_R(\ve p)]}}\,\right]^2.%
\end{split}\end{equation}
The constant $E^{(2)}$ is the two-nucleon separation energy averaged
over configurations of the $(A-2)$ nucleon system with low
excitation energy.

Even in nuclei with equal number of neutrons~$N$ and protons~$Z$
there is a~nonzero isovector contribution coming from Coulomb
interaction and isospin-dependent effects in other interactions, but
the standard approach is to neglect differences between the neutron
and proton SF. Taking into account differences between the neutron
and proton energy levels and their occupation probabilities in the
approach presented here yields only small change of the threshold
energy~$E^{(1)}$, so it only slightly shifts the plot of the
differential cross section. The reason of using the proton momentum
distribution in both cases is purely practical: it's known with
better accuracy.

For nonsymmetric nuclei, such as argon $^{40}_{18}$Ar, the neutron
SF must be different than the proton one. Due to the Pauli
exclusion, the surplus neutrons can occupy only the Fermi level and
thus it is assumed that the dominant contribution
to~$P_\text{vec}(\ve p,E)$ comes from momenta and energies very
close to the Fermi level. Kulagin and Petti~\cite{ref:Kulagin&Petti}
propose the approximation
\begin{equation}\label{eq:isovectorSF}
P_\text{vec}(\ve p,E)=\frac{N-Z}2\frac1{4\pi \overline
p_F^2}\:\delta(\n p-\overline p_F)\delta(E-E_F).
\end{equation}

\section{Results}\label{sec:Results}
\subsection{Verification of the method - an example of oxygen}\label{sec:Oxygen}
It is important to check how good the presented approximation is for
oxygen, for which detailed computation of the spectral function
exists \cite{ref:Benhar&Farina&Nakamura}. To perform the comparison
we have to calculate the values of the constants appearing
in~\eqref{eq:meanFieldSF} and~\eqref{eq:correlationSF}: $E^{(1)}$,
$E^{(2)}$, and $\langle \ve p^2\rangle$.

One-nucleon separation energy $E^{(1)}$ has been introduced
in~\eqref{eq:meanFieldSF} as an average energy of the
single-particle states below the Fermi level, so
\begin{eqnarray}\label{eq:E^(1)}
E^{(1)}&=&\frac{1}A\sum_{\alpha<\alpha_F}A_{\alpha}E_{\alpha}\int
n_{\alpha}(\ve
p)\:d^3p = \frac1A\sum_{\alpha<\alpha_F}A_{\alpha}E_{\alpha}c_{\alpha},\nonumber\\
\end{eqnarray}
where $c_{\alpha}$ is the occupation probability of the
state~$\alpha$.

\begin{table}
\caption{\label{tab:OccProbs}Occupation probabilities
$c_{nl}$~\cite{ref:OccProbs} and numbers of particles
$Z_{nl}=\frac12A_{nl}$ for proton shell-model states in $^{16}$O.}
\begin{ruledtabular}
\begin{tabular}{ldd}
State & \multicolumn{1}{c}{\mbox{$^{16}$O}} & \multicolumn{1}{c}{\mbox{$Z_{nl}$}}\\
\hline
$2s$ &0.06 & 2\\
$1d$ &0.14 & 10\\
$\alpha_F$& &\\
$1p$ &0.80 &6 \\
$1s$ &0.87 &2 \\
\end{tabular}
\end{ruledtabular}
\end{table}
\begin{table}
\caption{\label{tab:OxyEnLevels}Energy spectrum of protons and
neutrons in $^{16}$O~\cite{ref:OxyEnLevels} (the data for
$1s_{1/2}$ and $\alpha_F$ are taken
from~\cite{ref:Bohr&Mottelson}).}
\begin{ruledtabular}
\begin{tabular}{ldd}
State $\alpha$& \multicolumn{1}{c}{\mbox{Protons}} & \multicolumn{1}{c}{\mbox{Neutrons}}\\
\hline
$1d_{3/2}$ & -4.65 &  -0.93\\
$2s_{1/2}$ & 0.08& 3.27\\
$1d_{5/2}$ & 0.59& 4.15\\
$\alpha_F$& & 10\\
$1p_{1/2}$ & 12.11& 15.65\\
$1p_{3/2}$ & 18.44 &21.8\\
$1s_{1/2}$ & 45   & 47  \\
\end{tabular}
\end{ruledtabular}
\end{table}

Multi-index $\alpha$ is a compact way of writing quantum numbers
$(nlj)$; therefore $A_\alpha\equiv A_{nlj}$, etc. The occupation
probabilities $c_{nlj}$ for oxygen $^{16}$O were not available to
us, but we found values of $c_{nl}$ defined as
\begin{eqnarray}
c_{nl}&=&\frac{2(l+\frac12)+1}{2(2l+1)}\:c_{nl(l+\frac12)}+\frac{2(l-\frac12)+1}{2(2l+1)}\:c_{nl(l-\frac12)}\nonumber\\
&=&\frac{l+1}{2l+1}\:c_{nl(l+\frac12)}+\frac{l}{2l+1}\:c_{nl(l-\frac12)}.
\end{eqnarray}
It follows from the data for calcium in Sec.~\ref{sec:Argon} that
the values of $c_{nlj}$ for the levels differing only in spin
state are very similar. Assuming that $c_{nl(l+\frac12)}\approx
c_{nl(l-\frac12)}$ (or equivalently that $c_{nl(l-\frac12)}\approx
c_{nl}$ and $c_{nl(l+\frac12)}\approx c_{nl}$) and using data from
Tables~\ref{tab:OccProbs} and~\ref{tab:OxyEnLevels} we get the
following values:
\[\begin{split}
E^{(1)}_{^{16}\text{O}}&=19.18\text{ MeV for protons,}\\
E^{(1)}_{^{16}\text{O}}&=22.32\text{ MeV for neutrons.}
\end{split}\]
The Benhar's spectral function was calculated for protons;
therefore, to make appropriate comparisons, we used
$E^{(1)}_{^{16}\text{O}}=19.18$ MeV in our numerical calculations.
We checked also that applying the neutron value does not change the
cross sections significantly.

Two-nucleon separation energy $E^{(2)}$ can be approximated
by~\cite{ref:Kulagin&Petti}
\begin{eqnarray}\label{eq:E^(2)}
E^{(2)}=M_{A-2}+2M-M_A,
\end{eqnarray}
and for oxygen we obtained
\[
E^{(2)}_{^{16}\text{O}}=26.33~\text {MeV}.
\]
The mean value of $\ve p^2$ is
\begin{eqnarray}
\langle \ve p^2\rangle%
\equiv\frac{\int \ve p^2 n_\text{MF}(\ve p)\:d^3p}{\int n_\text{MF}(\ve p)\:d^3p}.%
\end{eqnarray}
Using the mean field momentum distribution for $^{16}$O as
in~\cite{ref:Ciofi&Simula}, we obtain
\[
\langle \ve p^2\rangle_{^{16}\text{O}}=\frac34(\hbar c)^2 0.9 \text{
fm}^{-2}=(162.2\text{ MeV}/c)^2.
\]

\begin{figure}
\graphicspath{{plot/}}
\includegraphics[width=0.48\textwidth]{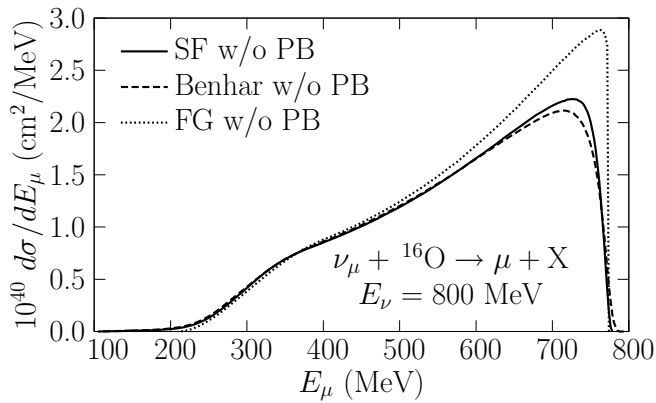}%
\caption{\label{fig:Oxy} Comparison of the quasielastic differential
cross section $d\sigma/dE_\mu$ of $^{16}$O for the Benhar's spectral
function (dashed), our spectral function (solid), and the Fermi gas
model with $p_F=225$ MeV (dotted line). Pauli blocking is not
included.}
\end{figure}
\begin{figure}
\graphicspath{{plot/}}
\includegraphics[width=0.48\textwidth]{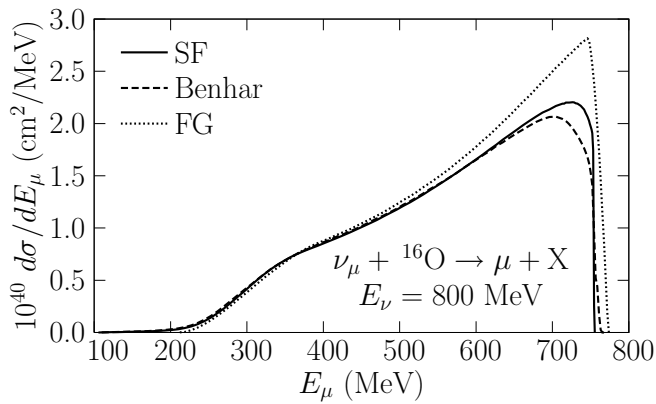}%
\caption{\label{fig:Oxyb} Same as Fig.~\ref{fig:Oxy}, but with Pauli
blocking (details in the text).}
\end{figure}
\begin{figure}
\graphicspath{{plot/}}
\includegraphics[width=0.48\textwidth]{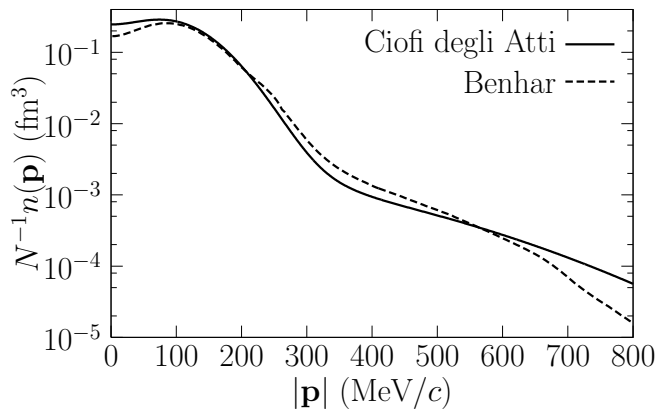}%
\caption{\label{fig:nBC} Comparison of the nucleon momentum
distribution computed from the Benhar's spectral function of
$^{16}$O and the distribution given by Ciofi degli Atti and
Simula~\cite{ref:Ciofi&Simula} used in our computations.}
\end{figure}
\begin{figure}
\graphicspath{{plot/}}
\includegraphics[width=0.48\textwidth]{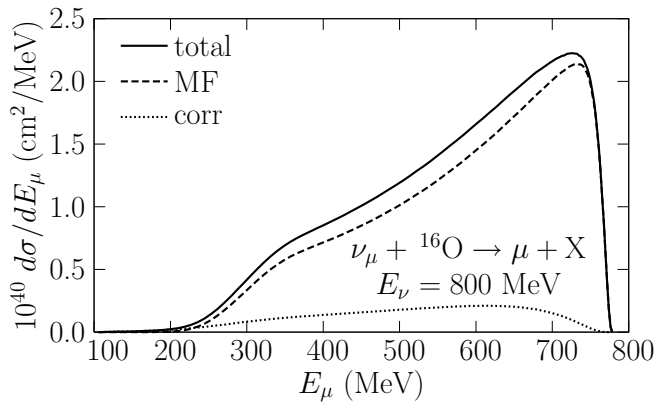}%
\caption{\label{fig:csContribs} Contributions to the quasielastic
differential cross section $d\sigma/dE_\mu$ of $^{16}$O in our
spectral function with Pauli blocking ($\overline p_F=209$ MeV).
Mean field and correlation contribution are shown by dashed and
dotted line, respectively. The solid line represents their sum.}
\end{figure}

Figure~\ref{fig:Oxy} shows a comparison of predictions for the
oxygen differential cross section in energy transfer given by our
spectral function with predictions given by the Benhar's spectral
function and the Fermi gas model. All numerical results in this
paper have been obtained using the dipole parameterization of the
form factors and axial mass $M_A = 1.03$ GeV.

To make comparisons with other numerical computations, we included
the Pauli blocking effect, taking the \emph{average} Fermi momentum
calculated from the known density profile of a~nucleus. The values
we use are $\overline p_F=209$~MeV \cite{ref:Benhar&Farina&Nakamura}
and $\overline p_F=217$ MeV~\cite{ref:Juszczak&Nowak&Sobczyk} for
oxygen and argon, respectively. Note that the corresponding values
needed for the FG model to reproduce the quasielastic peak in
electron scattering~\cite{ref:pF} (namely, $p_F=225$~MeV and
$p_F=251$~MeV) are visibly higher. We checked that the use of the
same value for Pauli blocking in both FG and SF does not reduce
discrepancy between them significantly.

The difference between predictions of two spectral functions is not
large, and it is greater if the Pauli blocking is included, see
Fig~\ref{fig:Oxyb}. We identify two reasons why the discrepancy is
observed. A part of it is explained by the different momentum
distributions in the spectral functions (see Fig.~\ref{fig:nBC}). We
deduce that the difference must be also caused by oversimplified
treatment of the $P_\text{MF}(\ve p,E)$, because for high $E_\mu$
this part gives dominant contribution to the differential cross
section, as shown in Fig.~\ref{fig:csContribs}.

In spite of simplicity of our approach, the results are in
satisfactory agreement with the ``exact'' spectral function, and we
continue our investigation to obtain a~description of the argon
nucleus.

\begin{table}[h]
\caption{\label{tab:CaProtons}Energy $E_\alpha$, occupation
probability $c_\alpha$, and number of particles
$Z_\alpha=\frac12A_\alpha$ for proton shell-model states in
$^{40}_{20}$Ca~\cite{ref:CaProtonData}.}
\begin{ruledtabular}
\begin{tabular}{lddd}
\multicolumn{1}{c}{$\alpha$}& \multicolumn{1}{c}{\mbox{$E_\alpha$}} & \multicolumn{1}{c}{\mbox{$c_\alpha$}} & \multicolumn{1}{c}{\mbox{$Z_\alpha$}}\\
\hline
$1g_{7/2}$&-21.00&0.04&8\\
$1g_{9/2}$&-9.75&0.06&10\\
$1f_{5/2}$&-4.51&0.08&6\\
$2p_{1/2}$&-2.02&0.07&2\\
$2p_{3/2}$&-0.92&0.08&4\\
$1f_{7/2}$&1.15&0.14&8\\
$\alpha_F$& 4.71&  &\\
$1d_{3/2}$&8.88&0.85&4\\
$2s_{1/2}$&10.67&0.87&2\\
$1d_{5/2}$&14.95&0.88&6\\
$1p_{1/2}$&31.62&0.91&2\\
$1p_{3/2}$&36.52&0.92&4\\
$1s_{1/2}$&57.38&0.93&2\\
\end{tabular}
\end{ruledtabular}
\end{table}
\begin{table}[h]
\caption{\label{tab:CaNeutrons}Energy $E_\alpha$, occupation
probability $c_\alpha$, and number of particles
$N_\alpha=\frac12A_\alpha$ for neutron shell-model states in
$^{40}_{20}$Ca~\cite{ref:CaNeutronData}.}
\begin{ruledtabular}
\begin{tabular}{lddd}
\multicolumn{1}{c}{$\alpha$}& \multicolumn{1}{c}{\mbox{$E_\alpha$}} & \multicolumn{1}{c}{\mbox{$c_\alpha$}} & \multicolumn{1}{c}{\mbox{$N_\alpha$}}\\
\hline
$1f_{5/2}$&1.50&0.07&6\\
$2p_{1/2}$&4.19&0.07&2\\
$2p_{3/2}$&5.59&0.08&4\\
$1f_{7/2}$&8.54&0.12&8\\
$\alpha_F$& 12&  &\\
$1d_{3/2}$&15.79&0.89&4\\
$2s_{1/2}$&17.53&0.91&2\\
$1d_{5/2}$&22.48&0.93&6\\
$1p_{1/2}$&39.12&0.96&2\\
$1p_{3/2}$&43.8 &0.96&4\\
$1s_{1/2}$&66.12&0.97&2\\
\end{tabular}
\end{ruledtabular}
\end{table}

\subsection{Argon spectral function}\label{sec:Argon}
In nature, the most common isotope of argon is $^{40}_{18}$Ar. The
feature of this nucleus which was not present in the previous
calculations is its lack of isospin symmetry. It is obvious that for
such nuclei, the vector contribution to the spectral function cannot
be neglected, and therefore we estimate it using
Eq.~\eqref{eq:isovectorSF}.

\begin{figure}
\graphicspath{{plot/}}
\includegraphics[width=0.48\textwidth]{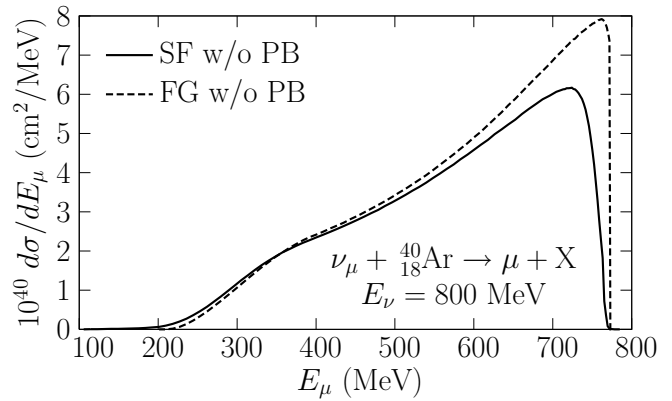}%
\caption{\label{fig:ArFG} Quasielastic differential cross section
$d\sigma/dE_\mu$ off $^{40}_{18}$Ar as a~function of produced muon
energy $E_\mu$ for the Fermi gas model ($p_F=251$ MeV, $\EB=28$ MeV)
and the spectral function. Pauli blocking is absent.}
\end{figure}
\begin{figure}
\graphicspath{{plot/}}
\includegraphics[width=0.48\textwidth]{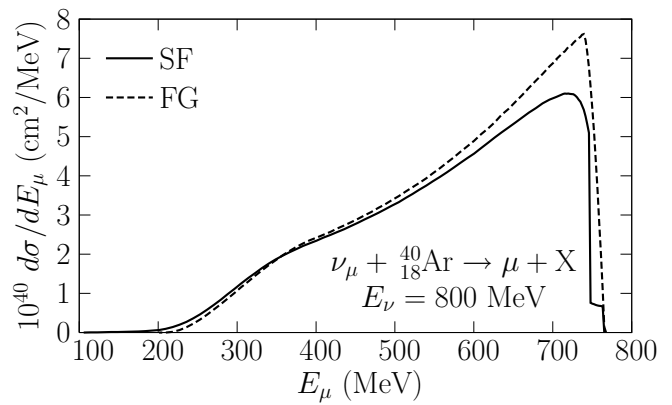}%
\caption{\label{fig:ArFGb} Same as Fig.~\ref{fig:ArFG}, but Pauli
blocking is present.}
\end{figure}
\begin{figure}
\graphicspath{{plot/}}
\includegraphics[width=0.48\textwidth]{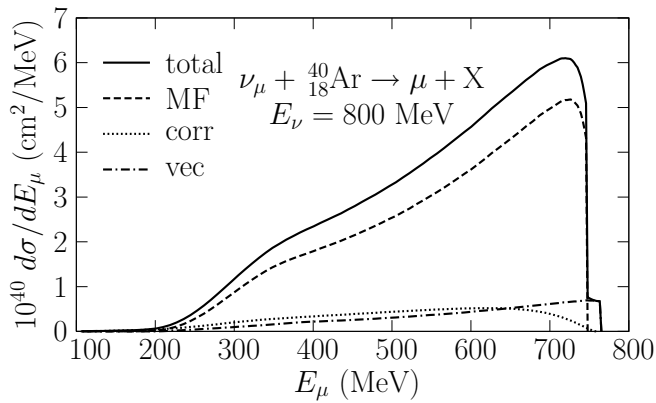}%
\caption{\label{fig:ArbContribs} Contributions to the differential
cross section shown in Fig.~\ref{fig:ArFGb} from $P_\text{MF}(\ve
p,E)$, $P_\text{corr}(\ve p,E)$ and $P_\text{vec}(\ve p,E)$. }
\end{figure}

We lack data on the energy levels and occupation probabilities for
the argon. The best we can do is to assume that they do not differ
significantly from those for calcium given in
Tables~\ref{tab:CaProtons} and \ref{tab:CaNeutrons} and use them to
obtain the value of one-nucleon separation energy. In this way we
get for neutrons
\[
E^{(1)}_{^{40}_{18}\text{Ar}}=29.26\text{ MeV.}
\]

From Eq.~\eqref{eq:E^(2)} two-nucleon separation energy for argon is
\[
E^{(2)}_{^{40}_{18}\text{Ar}}=16.46\text{ MeV.}
\]

We approximate the momentum distribution of argon with the one of
calcium~\cite{ref:Ciofi&Simula}, what yields
\[
\langle \ve p^2\rangle_{^{40}_{18}\text{Ar}}=\frac34(\hbar c)^2 1.08\text{ fm}^2=(177.2\text{ MeV})^2.%
\]

Using the argon spectral function, we calculated the inclusive cross
section for neutrino-argon interaction first without
(Fig.~\ref{fig:ArFG}) and then with the Pauli blocking
(Fig.~\ref{fig:ArFGb}). In Fig.~\ref{fig:ArbContribs}, we show the
contributions from $P_\text{vec}(\ve p,E)$, $P_\text{MF}(\ve p,E)$
and $P_\text{corr}(\ve p,E)$. We see that $P_\text{vec}(\ve p,E)$ is
clearly responsible for the singular behavior of the differential
cross section at small values of the energy transfer. The behavior
at the quasielastic peak is determined mainly by the
$P_\text{MF}(\ve p,E)$ and the detail knowledge of argon energy
levels can lead to modifications in this region. We also compared
the differential cross section per neutron for oxygen and argon and
found that they are rather similar.

\begin{figure}[b]
\graphicspath{{plot/}}
\includegraphics[width=0.48\textwidth]{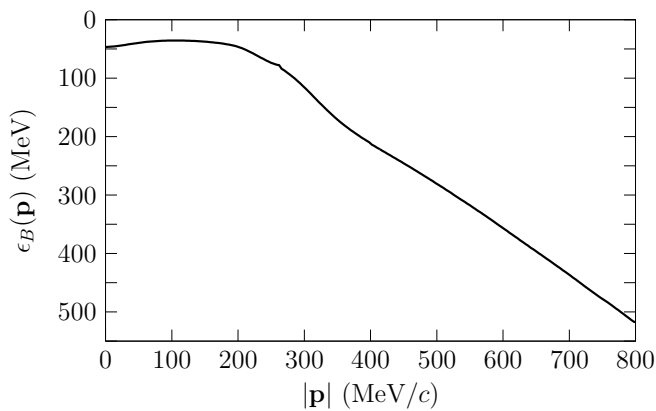}%
\caption{\label{fig:removE} Binding energy dependence on nucleon
momentum $\epsilon_B(\ve p)$ extracted from the Benhar's spectral
function.}
\end{figure}

\subsection{Effective description of nucleus}\label{sec:EffectiveDescription}%
Many Monte Carlo event generators rely on the Smith-Moniz version of
the FG model despite the fact that the energy conservation given by
Eq.~\eqref{ref:EB} and the step-function momentum distribution
provide a rather crude approximation of the real nucleus.

The use of the energy conservation written in~\eqref{ref:EB(p)} and
the momentum distribution of a specific nucleus does not complicate
model too much. This approach is simpler than the one based on the
complete spectral function and is easy to apply in MC simulations.
We refer to it as the \emph{effective description}. It is defined by
two functions $\epsilon_B(\ve p)$ and $n(\ve p)$ which characterize
specific nucleus and can be calculated directly from the spectral
function
\[
n(\ve p)={\int d E P(\ve p,E)}
\]
and
\begin{equation}
\epsilon_B(\ve p)=\frac1{n(\ve p)}\int d E \big(\sqrt{M^2+\ve p^2}-M+E\big)P(\ve p,E).%
\end{equation}
The momentum distribution $n(\ve p)$ obtained from the Benhar's
spectral function for oxygen was presented in Fig.~\ref{fig:nBC},
and the binding energy dependence on momentum is shown in
Fig.~\ref{fig:removE}. One can see that high-momentum nucleons are
deeply bound~\cite{ref:Ciofi&Simula&Frankfurt&Strikman} and they
cannot leave a nucleus. One can also see that for $\n p \in [0;200]$
MeV, the binding energy is roughly constant, varying between $\sim
35$ and $\sim 47$ MeV.

The cross section in the effective description is of the form
\begin{eqnarray}
\frac{d\sigma_\text{eff}}{d E_\mu}&=&\frac{G^2_F\cos^2\theta_C}{4\pi E_\nu^2}\frac{N}{\int d^3 p\:n(\ve p)}\int d{\n{q}}\,d^3p\:n(\ve p)\nonumber\\%
&&\times\delta\big(\omega+E_{\ve p}-\epsilon_B(\ve p)-E_\ve{p'}\big)\:\theta(\n{p+q}-\overline p_F)\nonumber\\%
&&\times\frac{\n q}{E_{\ve p}E_\ve{p'}}L_{\mu\nu}\widetilde H^{\mu\nu}_\text{eff}.%
\end{eqnarray}
The only difference between $L_{\mu\nu}\widetilde
H^{\mu\nu}_\text{eff}$ and $L_{\mu\nu}\widetilde
H^{\mu\nu}_\text{S-M}$ lies in the energy transfer, because now
$\widetilde\omega=\omega-\epsilon_B(\ve p)$.

\begin{figure}[t]
\graphicspath{{plot/}}
\includegraphics[width=0.48\textwidth]{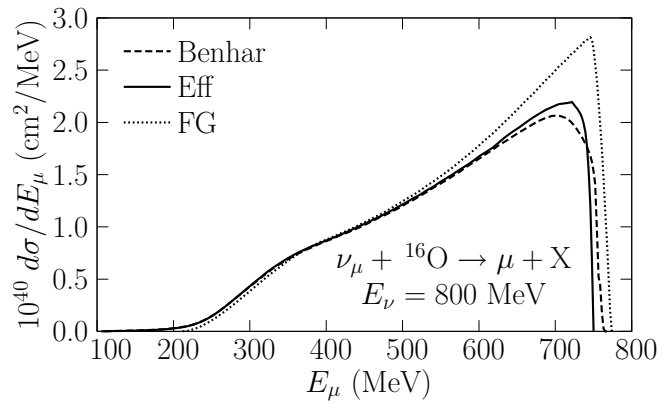}%
\caption{\label{fig:eff} Quasielastic differential cross section
$d\sigma/dE_\mu$ off $^{16}$O obtained from the effective
description, the Benhar's spectral function and the Fermi gas.
Pauli blocking is present.}
\end{figure}
\begin{figure}
\graphicspath{{plot/}}
\includegraphics[width=0.48\textwidth]{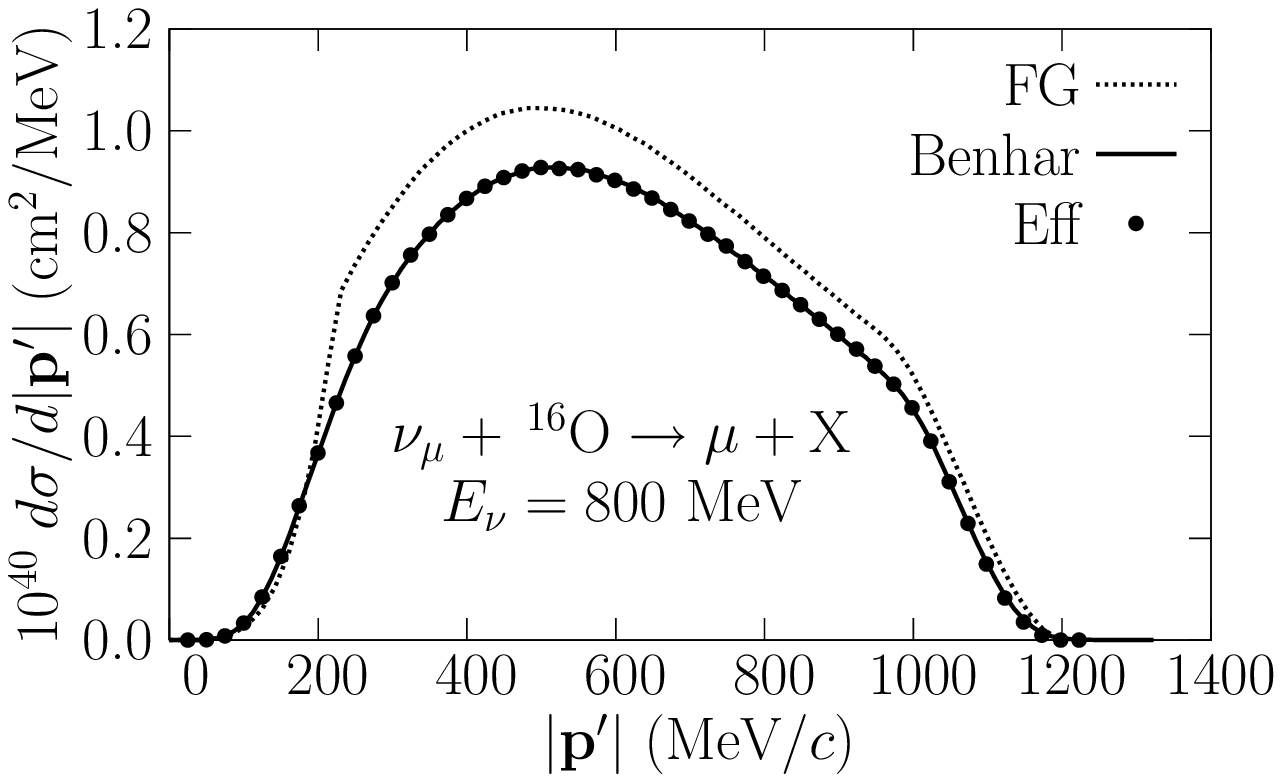}%
\caption{\label{fig:eff_h} Quasielastic differential cross section
$d\sigma/d\n{p'}$ off $^{16}$O obtained from the effective
description, the Benhar's spectral function and the Fermi gas
($p_F=225$ MeV, \EB=27 MeV). Pauli blocking is absent.}
\end{figure}
\begin{figure}
\graphicspath{{plot/}}
\includegraphics[width=0.48\textwidth]{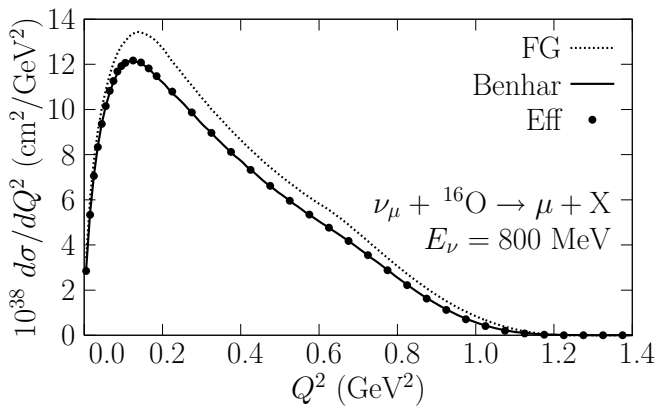}%
\caption{\label{fig:effQ2} Quasielastic differential cross section
in $Q^2=-q^2$ of $^{16}$O obtained from the effective description,
the Benhar's spectral function and the Fermi gas ($p_F=225$ MeV,
\EB=27 MeV). Pauli blocking is present.}
\end{figure}

\begin{figure}
\graphicspath{{plot/}}
\includegraphics[width=0.48\textwidth]{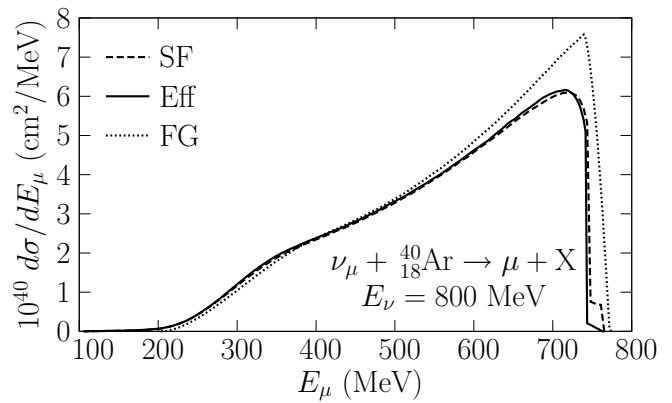}%
\caption{\label{fig:effAr} Quasielastic differential cross section
$d\sigma/dE_\mu$ of $_{18}^{40}$Ar obtained from the effective
description, our spectral function and the Fermi gas ($p_F=251$ MeV,
$\EB=28$ MeV). Pauli blocking is present.}
\end{figure}

\begin{figure}
\graphicspath{{plot/}}
\includegraphics[width=0.48\textwidth]{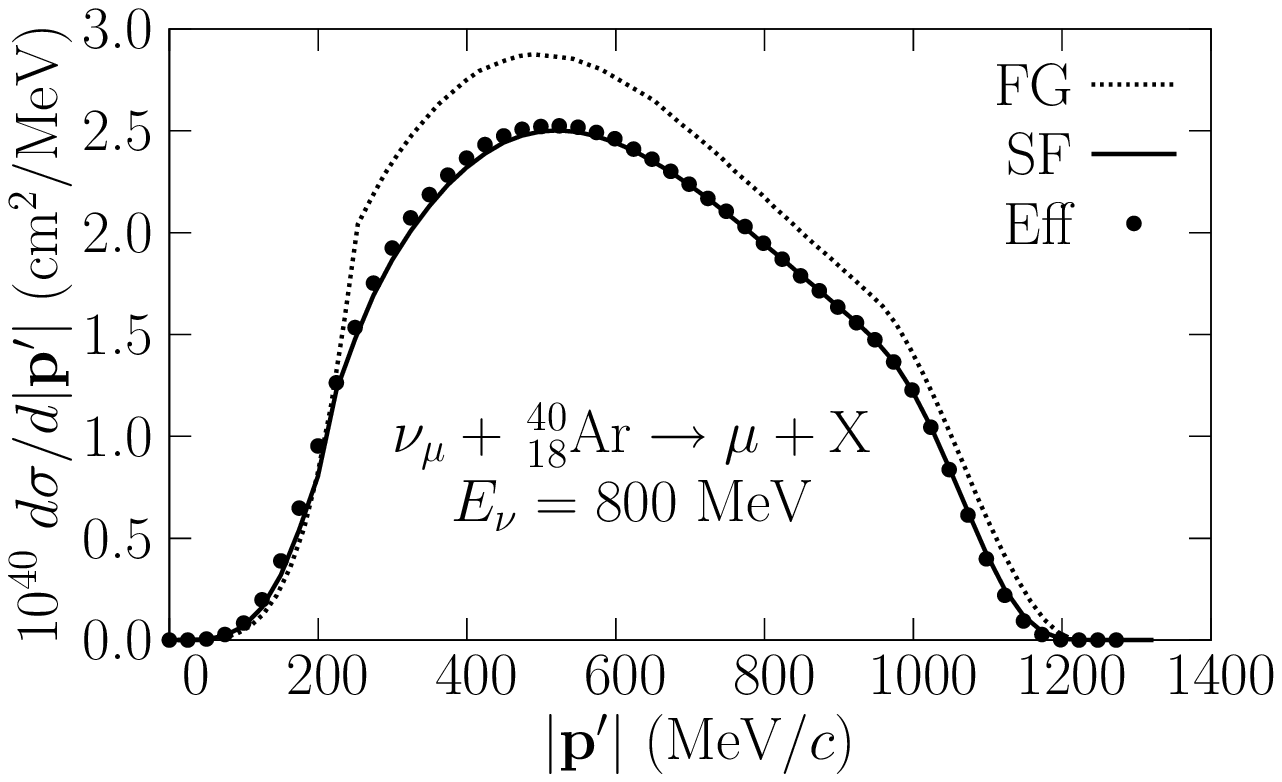}%
\caption{\label{fig:effArh} Quasielastic differential cross section
$d\sigma/d\n{p'}$ of $_{18}^{40}$Ar obtained from the effective
description, our spectral function and the Fermi gas ($p_F=251$ MeV,
$\EB=28$ MeV). Pauli blocking is absent.}
\end{figure}

Figs.~\ref{fig:eff}--\ref{fig:effArh} show results obtained within
this model.

The effective description of oxygen makes use of the Benhar's
spectral function. The cross section $d\sigma_\text{eff}/dE_\mu$
(Fig.~\ref{fig:eff}) is only slightly different with respect to the
one calculated with the spectral function. The difference is seen at
low energy transfer. We notice that the effective description
represents a major improvement over the FG model.

We compare also predictions of the three models for the distribution
of nucleons ejected from oxygen. The results are shown in
Fig.~\ref{fig:eff_h}. In the computations we did not apply the Pauli
blocking. Its straightforward implementation simply eliminates
nucleons with momenta lower then the Fermi momentum. Predictions of
the spectral function and the effective description are almost
identical. It is because the distributions of momenta of two
approaches are the same. We see again that the effective description
is a good approximation of the spectral function.

As is shown in Fig.~\ref{fig:effQ2}, the effective model also
succeeds in precisely reproducing the SF's differential cross
section $d\sigma/dQ^2$.


Finally  in Figs.~\ref{fig:effAr} and~\ref{fig:effArh} we show a
comparison of predictions from our spectral function of argon with
the effective description based on it. The differences between the
two models are smaller then in the case of oxygen because of the way
in which the dominant mean field part of the argon SF was treated.
We notice that the singularity present in Fig.~\ref{fig:effAr}
caused by the treatment of isovector part of the spectral function
is not seen in Fig~\ref{fig:effArh}.

\section{Conclusions}\label{sec:Con}%
We constructed the spectral function for argon and applied it in the
PWIA computation of the CC quasielastic cross section of
neutrino-argon scattering. The construction method was based on the
ideas contained in~\cite{ref:Kulagin&Petti}. We verified that when
applied to oxygen, the method leads to the results similar to those
obtained from the more detailed spectral function obtained by
Benhar~\cite{ref:Benhar&Farina&Nakamura}. The only FSI effect
included in the discussion was Pauli blocking. We proposed also the
effective approach in which the information contained in the
spectral function is used to define two functions. We argued that
this approach yields results similar to the spectral function, but
at the same time it can be easily applied in MC routines.

We see two possible applications of our results. The argon spectral
function can be used in more precise computations of the neutrino
cross sections in a few GeV energy region. The effective approach
may be used in existing MC event generators giving rise to the
important improvement with respect to the Fermi gas model. For
example, Fig.~\ref{fig:effQ2} shows that it reconstructs well the
$Q^2$ distribution of events which is important because of the
observed low $Q^2$ deficit of events in the K2K experiment. The
effective approach can be also used in cross section computations
for other targets important in neutrino experiments like carbon and
iron for which appropriate SFs exist \cite{ref:Benhar_priv}.

\begin{acknowledgments}
We thank Omar Benhar for providing us with the spectral function for
oxygen. One of the authors (JTS) was supported by the KBN grant
105/E-344/SPB/ICARUS/P-03/DZ211/2003-2005.
\end{acknowledgments}

\end{document}